\documentclass[11pt,a4paper]{article}

\usepackage[latin1] {inputenc}
\usepackage{amsmath}
\usepackage{amsfonts}
\usepackage{amssymb}
\usepackage{theorem}
\usepackage{amsopn}
\usepackage{calc}
\usepackage{fancybox}
\usepackage{cmmib57}

\DeclareMathOperator{\im}{Im}

\DeclareMathOperator{\re}{Re}

\theoremstyle{break}
\newtheorem{defin}{Definition}
\newtheorem{theo}{Theorem}
\newtheorem{prop}{Proposition}

\newtheorem{remark}{Remark}
\def\cqfd{ \hspace*{\fill} $ \Box $ \\ }
\def\proo{\noindent{\bf Proof: \\ }}

\def\CC{{\mathbb C}}

\def\LL{{\mathbb L}}

\def\RR{{\mathbb R}}

\def\ZZ{{\mathbb Z}}

\def\mcA{{\mathcal A}}

\def\mcD{{\mathcal D}}

\def\mcK{{\mathcal K}}
\def\mcH{{\mathcal H}}
\def\mcM{{\mathcal M}}

\def\mcT{{\mathcal T}}

\def\mcW{{\mathcal W}}

\def\mfh{{\mathfrak h}}

\def\O2n{\textup{\large{$ O $}}(2n)}

\def\Id{\textup{$ 1 \mspace{-6.8mu} \text{\large{1}} $}}

\def\tra{\textup{\textsf{\tiny{T}}}}

\def\ie{{\it i.e. }}

\begin{document}

\begin{center}
{\Large{\bf

Upper Quantum Lyapunov Exponent and  \\

\vspace{.15cm}

Anosov relations for quantum systems   \\

\vspace{.15cm}

driven by a classical flow \\

\vspace{1cm}}}

{\bf O. Sapin, H. R. Jauslin } \\
Laboratoire de Physique CNRS - UMR 5027 \\
Universit\'e de Bourgogne \\
BP 47870, F-21078 Dijon, France \\[0.3cm]
{\bf Stefan Weigert} \\
Department of Mathematics \\
University of York \\
Heslington YO10 5DD, UK

\end{center}

\vspace{4cm}

\begin{center}
{\bf Abstract}
\end{center}

We generalize the definition of quantum Anosov properties and the related
Lyapunov exponents to the case of quantum systems driven by a classical flow, \ie
skew-product systems. We show that the skew Anosov properties can be interpreted
as regular Anosov properties in an enlarged Hilbert space, in the framework of a
generalized Floquet theory. This extension allows us to describe the hyperbolicity
properties of almost-periodic quantum parametric oscillators and we show that their
upper Lyapunov exponents are positive and equal to the Lyapunov exponent of
the corresponding classical parametric oscillators. As second example, we show that the
configurational quantum cat system satisfies quantum Anosov properties.

\newpage

\section{Introduction}

Anosov properties and Lyapunov exponents are well-established characterization
of classical dynamics and it is natural to search for similar concepts applicable to
quantum dynamics. Several definitions have been given in the
literature (see \cite{MaKu,ViMe1,ViMe2,ENST,Thir,PeEm,Maje,MaVi1,Narn,MaVi2,JSGW}
and the references therein).  \\
In Ref. \cite{MaKu}, Majewski and Kuna defined a quantum Lyapunov exponent
for $ N $-level quantum systems. Later\footnote{Erratum to Ref. \cite{JSGW}: The
chronology of the Refs. \cite{MaKu} and \cite{ENST} as described in Refs.
\cite{JSGW} by two of the present authors is erroneous. To our knowledge the works
of \cite{MaKu} and \cite{ENST} were developped independently, while \cite{MaKu}
was published before \cite{ENST}.}, Emch, Narnhofer, Sewell and
Thirring \cite{ENST,Thir,Narn} proposed an axiomatic framework which allows one
to define an Anosov property for quantum mechanical systems. However,  the
resulting definition of a quantum Lyapunov exponent is limited since it only applies to
systems with a globally constant hyperbolicity property. \\
In Ref. \cite{JSGW}, the upper Lyapunov exponent for quantum systems in the
Heisenberg representation has been defined, close in spirit to definitions given in
Refs. \cite{MaKu,ENST}. Its usefulness has been illustrated with the example of
the parametric quantum oscillator with periodic time dependence. Moreover, it was
shown that whenever its upper Lyapunov exponent is positive, the system satisfies
the discrete quantum Anosov relations defined by Emch, Narnhofer, Sewell and
Thirring \cite{ENST,Thir,Narn}.  \\
In this paper we extend the study to systems described by a Hamiltonian operator
of the form $ H(\varphi^t (\theta)) $ (with $ \varphi^t $ a flow on a space $ \mcM $),
which will be referred to as {\it quantum skew-product system}. We generalize the
definition of Anosov relations so that it applies to this type of system. As in the case of
Floquet theory \cite{Howl,Howl2,Yaji1,JaLe,GeJa}, it is possible to make quantum
skew-product systems autonomous by embedding the dynamics in a larger Hilbert
space. The Anosov relations for quantum skew-product systems correspond
to the Anosov relations of the associated system in this enlarged Hilbert space. We
consider the parametric oscillator as an example. We show that the quantum parametric
oscillator verifies the Anosov relations for quantum skew-product systems if its upper
Lyapunov exponent is positive and the corresponding classical dynamics is reducible.
Thus the quantum parametric oscillator discussed in \cite{JSGW} is an Anosov quantum
skew-product system. As a second example we consider the configurational quantum cat
system \cite{Weig,Weig1,Weig2}, with periodic boundary conditions, which amounts to a
system with compact configuration space. \\
This paper is organized in the following way: In Section 2 we recall the
definition of the upper Lyapunov exponent and of the Anosov properties for a quantum
system describing the motion of a particle. In Section 3, we present the formalism of
quantum skew-product systems and the enlarged Hilbert space which allows one to turn
the system into an autonomous one. We propose a definition of the quantum Anosov
properties for quantum skew-product systems in Section 4. Finally, we treat the example
of the almost-periodic quantum parametric oscillator and the configurational quantum
cat system in Section 5.

\vspace{0.6cm}

\section{Upper Lyapunov exponents and quantum Anosov relations}

A quantum mechanical particle in one dimension is described by coordinate and
momentum operators $ \hat{x} $ and $ \hat{p} $ which satisfy the Heisenberg
commutation relation (we choose the units such that $ \hbar=1 $):
$$ [ \hat{x}, \hat{p} ] = i. $$
It is convenient to consider the $ C^* $-algebra generated by Weyl operators:
\begin{equation*}
 W (\beta, \gamma) \, = \, \exp \, [i \, (\beta \, \hat{x} + \gamma \, \hat{p})] ,
 \qquad \forall \, \beta, \gamma \in  \RR.
\end{equation*}
These operators satisfy the Weyl form of the commutation relations:
\begin{eqnarray*}
 W(\beta, \gamma)^{\dag} & = & W(-\beta,-\gamma), \\
 W(\beta, \gamma) \,W(\beta', \gamma') & = & e^{-\frac{i}{2}(\beta \, \gamma' \,
 - \, \gamma \, \beta')} \, W(\beta+\beta' , \gamma+\gamma').
\end{eqnarray*}
More abstractly, if the phase space is a real symplectic space $ V $ with
symplectic form $ \sigma $, the $ C^* $-algebra $ \mcW $ of canonical commutation
relations over $ (V,\sigma) $ is defined as the $ C^* $-algebra generated by
elements $ \{ \, W(\underline{\alpha}) \, | \, \underline{\alpha} \in V \} $ such that
\begin{eqnarray*}
 W(\underline{\alpha})^{\dag} & = & W(-\underline{\alpha}), \\
 W(\underline{\alpha}) \,W(\underline{\alpha}') & = & e^{-\frac{i}{2}
 \sigma(\underline{\alpha},\underline{\alpha}')} \, W(\underline{\alpha}
 +\underline{\alpha}').
\end{eqnarray*}
In this paper we consider only phase spaces $ V $ of finite dimension $ 2n $, with
the usual symplectic form
$$ \sigma(\underline{\alpha},\underline{\alpha}') \, = \, \alpha_x^{\tra} \,  \alpha_p'
\, - \, \alpha_p^{\tra} \, \alpha_x' \qquad \forall \, \underline{\alpha}= \left( \! \begin{array}{c}
\alpha_x \\ \alpha_p \end{array} \! \right), \underline{\alpha}' \in \RR^{2n}, $$
where $ \alpha_x^{\tra} $ denotes the transposed of $ \alpha_x $. Hence, the Weyl
operators can be written as:
\begin{equation*}
 W (\underline{\alpha}) \, = \, \exp \, [i \, (\alpha_x^{\tra} \, \hat{x} + \alpha_p^{\tra} \,
 \hat{p})] , \qquad  \underline{\alpha} \in  \RR^{2n}.
\end{equation*}
In order to define the quantum Lyapunov exponent, we consider derivations on
the algebra $ \mcW $. We denote by $ \delta_{\underline{\alpha}} $ the derivation
defined as the generator of the automorphism $ A \mapsto W(t \, \underline{\alpha})
\, A W(-t\, \underline{\alpha}) $ for all $ A \in \mcW $. Therefore we have
$$ \delta_{\underline{\alpha}}(A) \, \equiv  \,
                      [ \, L_{\underline{\alpha}} \, , \,  A \, ], \qquad \forall \, A \in \mcW, $$
where $ [\, ,\, ] $ is the commutator and
$$ L_{\underline{\alpha}} \, = \, \alpha_x^{\tra} \, \hat{x} \, + \, \alpha_p^{\tra} \, \hat{p},
 \qquad \underline{\alpha} \in  \RR^{2n}.  $$
In particular, we can check that
\begin{equation} \label{eq6}
 [ \, L_{\underline{\alpha}} \, , \, W(\underline{\alpha}')] \, = \, - \,
 \sigma(\underline{\alpha},\underline{\alpha}') \, W(\underline{\alpha}'),
 \qquad \forall \, \underline{\alpha},\underline{\alpha}' \in V.
\end{equation}

\noindent We assume that the dynamics defines an automorphism of $ \mcW $:
\begin{equation*}
 U^{\dagger}(t,t_0) \; A \; U(t,t_0) \, \equiv \, A(t,t_{0}) \in \mcW, ~~~~
 \forall \, A \in \mcW, \, \forall \, t ,t_0 \in  \RR,
\end{equation*}
where $ U(t,t_{0}) $ denotes the unitary propagator with initial time $ t_0 $. \\

\begin{defin}[cf. \cite{JSGW}]
 The {\it upper quantum Lyapunov exponent} is defined as
 \begin{equation*}
  \bar{\lambda} \, = \, \sup_{\underline{\alpha} \,  \in \, V} \, \bar{\lambda}_{\underline{\alpha}}
 \end{equation*}
 where
 \begin{equation*}
  \bar{\lambda}_{\underline{\alpha}} (U, L_{\underline{\alpha}}, A, t_0) \, := \,
  \limsup_{t \rightarrow \infty} \, \frac{1}{t} \, \ln \parallel
  [ \, L_{\underline{\alpha}} \, , \,  A(t, t_0) \, ] \parallel,
 \end{equation*}
 and $ \parallel . \parallel $ is the norm of the $ C^*$-algebra $ \mcW $.
\end{defin}

\noindent Since the time evolution is unitary, the exponent
$\bar{\lambda}_{\underline{\alpha}}$  can also be expressed as
\begin{equation}\label{lyaplt}
\bar{\lambda}_{\underline{\alpha}} (U, L_{\underline{\alpha}}, A, t_0) \,
= \, \limsup_{t \rightarrow \infty} \, \frac{1}{t} \, \ln \parallel [ \,
L_{\underline{\alpha}}(t_0,t) \, , \, A \, ] \parallel ,
\end{equation}
with
\begin{equation*}
L_{\underline{\alpha}}(t_0, t) \, := \,
                        U^\dagger(t_0, t) \; L_{\underline{\alpha}} \; U(t_0, t).
\end{equation*}

\begin{defin}
 A system satisfies the quantum Anosov relations \cite{ENST,Thir,Narn},
 if there are $ 2n $ directions $ \underline{\alpha}_1, \hdots, \underline{\alpha}_{2n}
 \! \in V $ such that the corresponding derivations satisfy for all $ t,t_0 \in \RR $
 $$  U(t,t_0) \; L_{\underline{\alpha}_i} \; U^{\dag}(t,t_0) \, = \,
                                               e^{\lambda_i (t-t_0)} \, L_{\underline{\alpha}_i}, $$
 where $ \lambda_{i} $ are $ 2n $ complex numbers such that
 $$ \re(\lambda_{1}) \, \leq \hdots  \leq \, \re(\lambda_{n}) \, < \, 0
 \, < \, \re(\lambda_{n+1}) \, \leq \hdots  \leq \, \re(\lambda_{2n}). $$
\end{defin}

\begin{remark}
 We have extend the definition of \cite{ENST} by allowing the numbers $ \lambda_{i} $
 to have an imaginary part. Moreover, we do not require that a state invariant
 under the actions of $ H $ and $ L_{\underline{\alpha}_i} $ exist.
\end{remark}

\begin{remark}
 To describe particles with internal structure, such as spin, it is necessary
 to generalize this construction. Assuming that the internal states of the particle
 form a complex Hilbert space $ \mfh $, the $ C^* $-algebra $ \mcW $ of canonical
 commutation relations over $ \mfh $ is by definition the $ C^* $-algebra generated
 by elements $ \{ \, W(\underline{\alpha}) \, | \, \underline{\alpha} \in \mfh \, \} $ such
 that
 \begin{eqnarray*}
  W(\underline{\alpha})^{\dag} & = & W(-\underline{\alpha}), \\
  W(\underline{\alpha}) \,W(\underline{\alpha}') & = &
  e^{-\frac{i}{2} \im(\langle \underline{\alpha},\underline{\alpha}' \rangle)} \,
  W(\underline{\alpha}+\underline{\alpha}'),
 \end{eqnarray*}
 where $ \langle . \, ,. \rangle $ denotes the scalar product in $ \mfh $.
\end{remark}

\vspace{0.6cm}

\section{Quantum skew-product systems and enlarged Hilbert space}

A quantum skew-product system is described by the following Schrödinger equation
with a non autonomous Hamiltonian in a Hilbert space $ \mcH $:
\begin{equation}  \label{eqSchrtheta}
 i \, \frac{d}{dt} \phi(t) \, = \, H(\varphi^{t}(\theta)) \, \phi(t),
\end{equation}
where $ \varphi^{t} $ is a continuous flow on a compact metric space
$ \mcM $ while $ H(\theta) $ is a self-adjoint operator depending
on the parameter $ \theta \in \mcM $ such that the evolution operator
$ U(t,t_{0}; \theta) $ exists and is strongly continuous with respect to
$ \theta \in \mcM $. This form of Hamiltonian operator includes periodic, quasi-periodic
and almost-periodic time dependence according to whether $ \mcM $ is a circle, a
torus or the hull of an almost-periodic function. \\
Any solution of (\ref{eqSchrtheta}) can be written as
$$ \phi(t;\theta) \, = \, U(t, t_{0}; \theta) \, \phi(t_{0}; \theta),  $$
with the operator $ U(t, t_{0}; \theta) $ satisfying
$$ i \frac{\partial}{\partial t}U(t, t_{0}; \theta) \, = \,
                                           H(\varphi^{t}(\theta)) \, U(t, t_{0}; \theta) $$
and $ U(t_{0}, t_{0};\theta)=\Id_{\mcH} $. \\
The uniqueness of solutions of (\ref{eqSchrtheta}) allows us to deduce the relations
\begin{eqnarray*}
 U(t, t_{1}; \theta) \; U(t_{1}, t_{0}; \theta) & = & U(t, t_{0}; \theta),   \\
 U(t+\tau, t_{0}+\tau; \theta) & = & U(t, t_{0}; \varphi^{\tau}(\theta)),
\end{eqnarray*}
for all $ t,t_{0},t_{1},\tau\in \RR $ and all $ \theta \in \mcM $.\\

\noindent Let $ \mu $ be an invariant probability measure on $ \mcM $. The
family of Koopman operators $ (\mcT^{t})_{_{t\in\RR}} $, defined by
$$ (\mcT^{t}\psi)(\theta) \, = \, \psi (\varphi^{t}(\theta))~~~~
                                   \text{for all} ~~ \psi \in \LL^{2}(\mcM,d\mu),  $$
is a strongly continuous one-parameter unitary group of operators. According to
Stone's theorem, there exists a self-adjoint operator $ G $ which is an infinitesimal
generator of $ \mcT^{t} $:
$$ \mcT^{t} \, = \, e^{itG} ~~~~ \text{for all} ~~ t \in \RR. $$

\noindent The separable Hilbert space $ \mcK=\LL^{2}(\mcM, d\mu; \mcH)=
\LL^{2}(\mcM,d\mu) \otimes \mcH $ will be called \emph{the enlarged space} of
$ \mcH $. The family of operators $ U(t,t_{0}; \theta) \in \mcH $ depending on the
parameter $ \theta \in \mcM $ defines a unitary operator acting in $ \mcK $ which
maps a function $ \theta \mapsto \psi(\theta)\in\mcH $ of $ \mcK $ to the function
$ \theta \mapsto U(t,t_{0}; \theta)\psi(\theta) \in\mcH $. To avoid a complicated
notation, we also denote this operator by $ U(t,t_{0}; \theta) $. Moreover, we omit the
identity factor of $ \mcT^{t}\otimes \Id_{\mcH} $ in the
Koopman operator in $ \mcK $. From the uniqueness of solutions of
(\ref{eqSchrtheta}) we can conclude that
$$ \mcT^{s} \, U(t,t_{0}; \theta) \, = \, U(t,t_{0}; \varphi^{s}(\theta)) \, \mcT^{s} $$
for all $ t,t_{0},s\in\RR $ and all $ \theta \in \mcM $.  \\

\begin{defin}
  We define a unitary operator $ U_{K}(t,t_{0}) $ acting on the enlarged
  space $ \mcK $ by
  $$ U_{K}(t,t_{0}) \, = \, \mcT^{-t} \, U(t,t_{0}; \theta) \, \mcT^{t_{0}}
                           \, = \, \mcT^{-(t-t_{0})} \, U(t-t_{0},0; \theta).           $$
\end{defin}

\vspace{0.1cm}

\noindent One can show that it is strongly continuous in $ t - t_{0} $, and
Stone's theorem implies that there is a self-adjoint operator $ K $ on $ \mcK $,
called {\it generalized Floquet Hamiltonian}, such that
$$ U_{K}(t,t_{0}) \, = \, e^{-i \, (t-t_{0}) K}. $$
The solution of the associated Schrödinger equation
\begin{equation} \label{eqSchrF}
   i \frac{d}{dt}\psi(t) \, = \, K \, \psi(t)
\end{equation}
reads $ \psi(t)=U_{K}(t,t_{0}) \, \psi(t_{0})\in \mcK $, and it is linked
to a solution $ \phi $ of the Schrödinger equation (\ref{eqSchrtheta})
in $ \mcH $ by
$$ \phi(t) \, = \, \mcT^{t} \, \psi(t) \, = \, \psi(t, \varphi^{t}(\theta)). $$

\begin{prop}
 We denote $ H(\theta) $ the self-adjoint operator on $ \mcK $ which maps
 $ \psi \in \mcK $ to the function $ \theta \mapsto H(\theta) \, \psi(\theta)
 \in \mcH $ of $ \mcK $. We assume that $ H(\theta) $ is a self-adjoint
 operator of $ \mcK $. We have the formal equality
 $$ K \, = \, G \, + \, H(\theta). $$
\end{prop}

\proo
The operator $ U_{K}(t,t_{0}) $ is strongly differentiable on $ \mcD(K) $,
and we can write formally
$$ i\frac{\partial}{\partial t}U_{K}(t,t_{0}) \, = \, K \, U_{K}(t,t_{0}) ~~~~
                                                               \text{for all} ~~ t,t_{0} \in \RR. $$
Therefore
\begin{eqnarray*}
 K & = & i\frac{\partial}{\partial t} U_{K}(t,t_{0})|_{t=t_{0}} \\
 & = & i\frac{\partial}{\partial t} \big(\mcT^{-(t-t_{0})} \,
 U(t-t_{0},0; \theta)\big) \big|_{t=t_{0}} \\
 & = & i\frac{\partial}{\partial t} \big(\mcT^{-(t-t_{0})}\big)\big|_{t=t_{0}} \,
 U(0,0;\theta) +i \,  \frac{\partial}{\partial t} U(t,t_{0};\theta) |_{t=t_{0}} \\
 & = & G \, + \, H(\theta).
\end{eqnarray*}
\cqfd

\vspace{0.6cm}

\section{Quantum skew-product Anosov properties}

For a quantum skew-product system defined by the Schrödinger equation
(\ref{eqSchrtheta}) with an Hamiltonian of the form
$ H(\hat{x},\hat{p},\varphi^{t}(\theta)) $,
we define the Anosov property by  \\
\begin{defin}
  A quantum skew-product system satisfies the quantum skew-product Anosov
  relations if there exist $ 2n $ functions
  $ \underline{\alpha}_1, \hdots, \underline{\alpha}_{2n} \! \! : \mcM \rightarrow V $
  such that the corresponding derivations satisfy for all $ t,t_0 \in \RR $ and
  $ \theta\in \mcM $
  $$ U(t,t_{0};\theta) \; L_{\underline{\alpha}_{i}(\varphi^{t_{0}}(\theta))}
                    \; U^{\dagger}(t,t_{0};\theta) \, = \,  e^{\lambda_{i}(t-t_{0})}
                                        \, L_{\underline{\alpha}_{i}(\varphi^{t}(\theta))}, $$
  where $ \lambda_{i} $ are $ 2n $ complex numbers  such that
  $$ \re(\lambda_{1}) \, \leq \hdots  \leq \, \re(\lambda_{n}) \, < \, 0
  \, < \, \re(\lambda_{n+1}) \, \leq \hdots  \leq \, \re(\lambda_{2n}). $$
\end{defin}

\vspace{0.3cm}

\noindent The operators $ L_{\underline{\alpha}_{i}(\theta)} $ define operators
$ L_{\underline{\alpha}_{i}} $ acting on the enlarged Hilbert space
$ \mcK=L^{2}(\mcM,\mu) \otimes \mcH $,
given by
$$ (L_{\underline{\alpha}_{i}} \psi )(\theta)=L_{\underline{\alpha}_{i}(\theta)}
\, \psi(\theta) ~~~~ \text{for all} ~~~  \psi\in \mcD(L_{\underline{\alpha}_{i}})
\subset \mcK. $$

\vspace{0.3cm}

\begin{prop}
 In the enlarged space $ \mcK=\LL^{2}(\mcM,\mu)\otimes \mcH $,
 the dynamics generated by the Hamiltonian $ K=G+H(\theta) $ satisfies
 the standard quantum Anosov properties \cite{ENST}:
 $$ U_{K}(t,t_{0}) \; L_{\underline{\alpha}_{i}} \; U_{K}^{\dagger}(t,t_{0})
                     \,  = \, e^{\lambda_{i}(t-t_{0})} \, L_{\underline{\alpha}_{i}}. $$
\end{prop}

\proo
By definition, the evolution operator satisfies
$$ U_{K}(t,t_{0})= \mcT^{-t} \, U(t,t_{0};\theta) \, \mcT^{t_{0}}. $$
Then the equation
$$ U_{K}(t,t_{0}) \; L_{\underline{\alpha}_{i}} \; U_{K}^{\dagger}(t,t_{0})
                        \, = \, e^{\lambda_{i}(t-t_{0})} \, L_{\underline{\alpha}_{i}} $$
can be written as
$$ \mcT^{-t} \, U(t,t_{0}) \, \mcT^{t_{0}} \;
     L_{\underline{\alpha}_{i}} \, \mcT^{-t_{0}} \; U(t,t_{0};\theta) \; \mcT^{t}
                        \,  = \, e^{\lambda_{i}(t-t_{0})} \, L_{\underline{\alpha}_{i}}. $$
Using the relation
$$ \mcT^{t} \, L_{\underline{\alpha}_{i}(\theta)} \,
            = \, \mcT^{t} \, \alpha_{i_{\scriptstyle x}} (\theta)^{\tra} \otimes \hat{x} \,
                   + \, \mcT^{t} \, \alpha_{i_{\scriptstyle p}} (\theta)^{\tra} \otimes \hat{p} \,
                          = \, L_{\underline{\alpha}_{i}(\varphi^{t}(\theta))} \, \mcT^{t} $$
for all $ t \in \RR $, we obtain
$$ U(t,t_{0};\theta) \; L_{\underline{\alpha}_{i}(\varphi^{t_{0}}(\theta))} \;
                              U^{\dagger}(t,t_{0};\theta) \, = \, e^{\lambda_{i}(t-t_{0})} \,
                                               L_{\underline{\alpha}_{i}(\varphi^{t}(\theta))}. $$
\cqfd

\vspace{0.6cm}

\section{Examples}

\subsection{The almost-periodic quantum parametric oscillator}

As in \cite{JSGW}, we consider the parametric quantum oscillator which is described
by the Hamiltonian (we take the mass = 1):
\begin{equation}\label{eq13}
H(t) \,  = \, \frac{1}{2} \, \hat{p}^2 \, + \, \frac{1}{2} \, f(t) \,  \hat{x}^2
\end{equation}
where $ f $ is an almost-periodic real valued function.  \\
The classical dynamics corresponding to the Hamiltonian (\ref{eq13}) has the same
form as the eigenvalue equation of the almost-periodic Schr\"odinger operator:
\begin{equation}\label{eq15}
-\ddot{x} \, + \, V(t) \, x = E \, x
\end{equation}
with $ f(t)=E- V(t) $. For a fixed almost-periodic real valued function $ V(t) $,
we will now analyze the one-parameter family of systems defined by varying
$ E $ on $ \CC $ and, in particular, when $ E $ is real and in the resolvent set
$ \rho $ of the almost-periodic Schr\"odinger operator $ -d^{2}/dt^{2}+ V(t)  $.  \\

\begin{theo}
For any observable $ A=W(\underline{\beta}) $ in the Weyl algebra, in
the instability region $ E  \in \rho \cap \RR $, there is a stable direction
$ \underline{\alpha}_s $, which depends on $ t_0 $, for which
\begin{equation*}
\bar{\lambda}_{\underline{\alpha}_s}(U,L_{\underline{\alpha}_s},A, t_0)
\, = \,  -\lambda_{c}<0,
\end{equation*}
whereas for all other directions $\underline{\alpha}$
\begin{equation*}
\bar{\lambda}_{\underline{\alpha}}(U,L_{\underline{\alpha}},A,t_0)
\, = \,  \lambda_{c}>0.
\end{equation*}
where $ \lambda_{c} $ is the Lyapunov exponent of the classical system.
Thus the upper quantum Lyapunov exponent is positive,
\begin{equation*}
\bar{\lambda} \, = \, \sup_{\underline{\alpha} }
\bar{\lambda}_{\underline{\alpha}} \, = \, \lambda_{c}>0.
\end{equation*}
\end{theo}

\proo
The spectral parameter $ E $ is in the resolvent set $ \rho $ of the operator if and only
if the classical system
\begin{equation}\label{sl2eq}
 \frac{d}{dt}  \left( \begin{array}{c} p \\ q \\ \end{array} \right)
 = \left( \begin{array}{cc} 0 & V(t)-E \\ 1 & 0 \\ \end{array} \right)
 \left( \begin{array}{c} p \\ q \\ \end{array} \right)
\end{equation}
has an exponential dichotomy \cite{John}. In particular, if $ E \in \rho $, the
system (\ref{eq15}) has two linearly independent solutions
$ q_{+}\in \LL^{2}([0,+\infty[) $ and $ q_{-}\in \LL^{2}(]-\infty,0]) $. \\
The functions
$$ m_\pm=\dfrac{p_\pm}{q_\pm} ~~~~ \text{ and } ~~~~
                                      \tilde{m}_\pm=\dfrac{p_\pm}{q_\pm+i \, p_\pm} ~ , $$
defined for $ E \notin \RR $ and  $ E \in \rho \cap \RR $, respectively, are
almost-periodic \cite{Scha,JoMo}.  \\

\noindent The classical Lyapunov exponent associated with the dynamics of
(\ref{eq15}) is defined as
$$ \lambda_{c} = \sup \big( \limsup_{t \rightarrow +\infty}
                                                              \dfrac{1}{2t} \ln (|p|^2+|q|^2)\big), $$
where the supremum is taken over all non trivial solutions $ (p, q) $ of (\ref{sl2eq}),
and it satisfies \cite{Joh3}
\begin{eqnarray} \label{explyapcla}
\lambda_{c} & = & - \limsup_{t\rightarrow +\infty} \dfrac{1}{2t} \ln (|p_+|^2+|q_+|^2)
\nonumber \\
& = & \limsup_{t\rightarrow +\infty} \dfrac{1}{2t} \ln (|p_-|^2+|q_-|^2).
\end{eqnarray}

\vspace{0.2cm}

In order to determine the upper quantum Lyapunov exponent, we first need to
calculate $ L_{\underline{\alpha}}(t_0,t) $ which we write in the form
\begin{equation}\label{eq17}
L_{\underline{\alpha}} (t,t_0) \, = \, \alpha_x (t,t_0) \, \hat{x} \, +
\, \alpha_p(t,t_0) \, \hat{p}.
\end{equation}

\noindent The propagator $ F(t,t_0) $ of the classical equation (\ref{sl2eq}),
defined by
\begin{equation*}
\left( \begin{array}{c} p(t) \\ x(t) \\ \end{array} \right)
= \, F(t, t_0) \left( \begin{array}{c} p(t_0) \\ x(t_0) \\
\end{array} \right), \qquad F(t, t) =1 \quad \forall t,
\end{equation*}
may be written as
\begin{equation} \label{propag}
F(t, t_0) \, = \, P(t) \left( \begin{array}{cc} \frac{\psi_+(t)}{\psi_+(t_0)} & 0 \\
0 & \frac{\psi_-(t)}{\psi_-(t_0)} \end{array} \right) P(t_{0})^{-1}
\end{equation}
where $ \psi_\pm(t)=q_\pm(t)+i \, p_\pm(t) $ and
$$ P(t)\, = \, \left( \begin{array}{cc}  \tilde{m}_+(t) & \tilde{m}_-(t) \\
            1-i \, \tilde{m}_+(t)  & 1-i \, \tilde{m}_-(t)  \end{array} \right).  $$

\noindent Using the fact that the  Heisenberg equations of motion for the
operators  $ \hat{x}(t) $ and $ \hat{p}(t) $ have the same form as the
classical equations for $ x(t) $ and $ p(t) $, we can write
$$ \left( \begin{array}{c} U^\dagger (t, t_0) \, \hat{p} \, U(t,t_0) \\
            U^\dagger (t, t_0) \, \hat{x} \, U(t, t_0) \end{array} \right) =
              F(t, t_0) \left( \begin{array}{c} \hat{p} \\ \hat{x} \end{array} \right) $$
Thus, using the relation
$$ L_{\underline{\alpha}}(t,t_0)=
\left( \begin{array}{c} \alpha_p \\ \alpha_x \end{array} \right)^\tra
\left( \begin{array}{c} U^\dagger (t, t_0) \, \hat{p} \, U(t, t_0) \\
U^\dagger (t, t_0) \, \hat{x} \, U(t, t_0) \end{array} \right) =
\left( \begin{array}{c} \alpha_p(t, t_0) \\ \alpha_x(t, t_0) \end{array} \right)^\tra
\left( \begin{array}{c} \hat{p} \\ \hat{x} \end{array} \right),  $$
we obtain
\begin{equation}  \label{alphtt0}
\left( \begin{array}{c} \alpha_p(t, t_0) \\ \alpha_x(t, t_0) \end{array} \right) = \,
\left( P(t_{0})^{-1} \right)^{\tra} \left( \begin{array}{cc} \frac{\psi_+(t)}{\psi_+(t_0)} & 0 \\
0 & \frac{\psi_-(t)}{\psi_-(t_0)} \end{array} \right) P(t)^{\tra}
\left( \begin{array}{c} \alpha_p \\ \alpha_x \end{array} \right).
\end{equation}

\vspace{0.3cm}

\noindent If $ A=W(\underline{\beta})=e^{i(\beta_x \hat{x}+\beta_p \hat{p})} $ then,
according to (\ref{eq6}),
$$ \big[ L_{\underline{\alpha}}(t_0, t)\, , \, A \big] \,  = \,
     \big( \alpha_{p}(t_0, t) \, \beta_x - \alpha_{x}(t_0, t) \, \beta_p \big) \, A \, = \,
                  -\sigma \big( \underline{\alpha}(t_0, t) \, , \, \underline{\beta} \big) \, A, $$
implying that
\begin{equation*}
\big{\Vert} \big[ L_{\underline{\alpha}}(t_0, t)\, , \, A \big] \big{\Vert} \, = \, \big|
\, \alpha_{p}(t_0, t) \, \beta_x - \alpha_{x}(t_0, t) \, \beta_p \, \big| \, = \, \big|
\sigma \big( \, \underline{\alpha}(t_0, t) \, , \, \underline{\beta} \big) \, \big|,
\end{equation*}
where we have used  $ \parallel A \parallel \, = \, 1 $.
By (\ref{alphtt0}), the stable direction $ \underline{\alpha}_s $
is given by
\begin{equation*}
\left( \begin{array}{c} \alpha_{ps} \\ \alpha_{xs} \end{array} \right)=
\left( \begin{array}{c} -q_+(t_0) \\ p_+(t_0) \end{array}
\right)\in \, \RR^2.
\end{equation*}
Indeed we obtain
\begin{equation*}
\left( \begin{array}{c} \alpha_{ps}(t_0, t) \\ \alpha_{xs}(t_0, t) \end{array}
\right)=\,\psi_+(t) \left( \begin{array}{c} -1+i \, \tilde{m}_+(t) \\ \tilde{m}_+(t)
\end{array} \right),
\end{equation*}
and
\begin{equation*}
\big{\Vert} \big[ L_{\underline{\alpha}_s}(t_0,t), A \big] \big{\Vert} = \big|
(1+i \, \tilde{m}_+(t)) \, \beta_x \, + \, \tilde{m}_+(t) \, \beta_p  \big| \, |
\psi_+(t) |.
\end{equation*}
According to (\ref{explyapcla}), the quantum Lyapunov exponent in this direction is
\begin{eqnarray*}
\lambda_{\underline{\alpha}_s}(U,L_{\underline{\alpha}_s}, A, t_0)
& = &  \limsup_{t\rightarrow +\infty} \dfrac{1}{t} \ln (|\psi_+(t)|) \\
& = &  \limsup_{t\rightarrow +\infty} \dfrac{1}{2t} \ln (|p_+(t)|^2+|q_+(t)|^2) \\
& = &  -\lambda_{c} \, < \, 0.
\end{eqnarray*}
For all other directions $ \underline{\alpha} \in \RR^2 $, it is easy to check that
the upper Lyapunov exponent is positive,
\begin{eqnarray*}
\lambda_{\underline{\alpha}}(U,L_{\underline{\alpha}},A, t_0)
& = &  -\limsup_{t\rightarrow +\infty} \dfrac{1}{t} \ln (|\psi_+(t)|) \\
& = &  -\limsup_{t\rightarrow +\infty} \dfrac{1}{2t} \ln (|p_+(t)|^2+|q_+(t)|^2) \\
& = &  \lambda_{c} \, > \, 0.
\end{eqnarray*}
\cqfd

\begin{remark}
The result of Theorem 1 can be extend to the multidimensional case where
$$  H(t) \,  = \, \frac{1}{2} \, \hat{p}^2 \,
                                      + \, \frac{1}{2} \, \hat{x}^{\tra} \, A(t) \,  \hat{x} $$
with $ A(t) $ a real symmetric matrix depending almost-periodically on time.
Writing $ A(t) =E \, \Id+V(t) $, the equations of motion of corresponding classical
system have the same form as the eigenvalue equation of the Schr\"odinger operator
$ -d^{2}/dt^{2}+ V(t)  $. In the instability region $ E  \in \rho \cap \RR $, there will be
$ n $ stable directions $ \underline{\alpha}_{s_{\scriptstyle i}} $, depending on
$ t_0 $, with negative Lyapunov exponent, while they will de positive for the remaining
directions.  The main argument is again the exponential dichotomy in the resolvent set
\cite{Joh2,JoNe}.
\end{remark}

\vspace{0.7 cm}

To study of the Anosov properties for the almost-periodic quantum parametric
oscillator, we formulate it as a quantum skew-product system,
\begin{equation}  \label{hamiltphase}
H(\varphi^{t}(\theta)) \, = \, \frac{1}{2} \, \hat{p}^2 \,
                             + \, \frac{1}{2} \, \tilde{f}(\varphi^{t}(\theta)) \, \hat{x}^2
\end{equation}
where $ \tilde{f} $ is the extension of the almost-periodic function $ f $ to a
continuous function on its hull and $ \varphi^{t} $ the associated minimal flow
(see \cite{JoMo}).\
As before we introduce a parameter $ E $ by writing
$ \tilde{f}(\varphi^{t}(\theta)) = E - V(\varphi^{t}(\theta)) $, and we denote
the hull of the almost-periodic function by $ \mcM $. \\
The corresponding classical system is now given by
\begin{equation}\label{eqclass}
 \frac{d}{dt}  \left( \begin{array}{c} p \\ x \\ \end{array} \right)
 = \left( \begin{array}{cc} 0 & E-V(\varphi^{t}(\theta)) \\ 1 & 0 \\ \end{array} \right)
 \left( \begin{array}{c} p \\ x \\ \end{array} \right).
\end{equation}

\begin{defin}
 A linear system of differential equations
 $$ y'(t) \, = \, A\big(\varphi^{t}(\theta)\big) \, y(t) $$
 with $ y(t)\in \RR^{n} $ and $ A(\theta) $ a matrix depending on $ \theta\in\mcM $,
 is called reducible if it can be transformed into a system with constant coefficients
 $$ z' \, = \, C \, z $$
 by a transformation $ y(t)=T\big(\varphi^{t}(\theta)\big) \, z(t) $ where $ T(\theta) $
 is a non singular matrix for all $ \theta\in\mcM $.
\end{defin}

\begin{remark}
 The system (\ref{eqclass}) is reducible when the potential is
 quasi-periodic with frequencies satisfying a Diophantine condition
 (\cite{JoSe,SaS3,MoPo}).
\end{remark}

\begin{theo}
  Let the classical system (\ref{eqclass}) be reducible. Then the corresponding
  quantum parametric oscillator satisfies the quantum skew-product Anosov
  properties for $ E $ being in the resolvent set, $ E\in \rho $: there exist two
  measurable functions
  $ \underline{\alpha}_{\pm}: \mcM \rightarrow \RR^{2} $ and $ \lambda_{\pm} $
  such that $ \pm \re(\lambda_{\pm})>0 $ and
  $$ U(t,t_{0};\theta) \, L_{\underline{\alpha}_{\pm}(\varphi^{t_{0}}(\theta))} \,
  U^{\dagger}(t,t_{0};\theta) \, = \, e^{\lambda_{\pm}(t-t_{0})} \,
  L_{\underline{\alpha}_{\pm}(\varphi^{t}(\theta))}, $$
  with $ L_{\underline{\alpha}_{\pm}(\theta)} \, = \, \alpha_{x\pm}(\theta) \, \hat{x}
  \, + \, \alpha_{p \pm}(\theta) \, \hat{p} $.
\end{theo}

\proo
Using reducibility and the hyperbolic character of the flow of (\ref{eqclass})
in the resolvent set , we obtain
\begin{equation*}
F(t,t_0;\theta) \, = \, g(\varphi^{t}(\theta)) \, \exp\left[(t - t_{0}) \left(
\begin{array}{cc} \lambda_{+} & 0 \\ 0 & - \lambda_{+} \\
\end{array} \right) \right] \, g(\varphi^{t_{0}}(\theta))^{-1}
\end{equation*}
where $ g $ is a is a non singular matrix for all $ \theta\in\mcM $ and
$ \re(\lambda_{+}) \geq  0 $.  \\
Consequently,
$$ \begin{array}{l}
\hspace{-0.2cm} \left( \begin{array}{c}  U^{\dagger}(t,t_{0};\theta) \, \hat{p}
\, U(t,t_{0};\theta) \\ U^{\dagger}(t,t_{0};\theta) \, \hat{x} \, U(t,t_{0};\theta)
\end{array} \right) = \, F(t,t_0;\theta) \left( \begin{array}{c} \hat{p} \\ \hat{x}
\end{array} \right) \\ \hspace{2.8cm} = \, g(\varphi^{t}(\theta)) \left(
\begin{array}{cc} e^{(t-t_{0}) \lambda_{+}} & 0 \\ 0 & e^{-(t-t_{0})
\lambda_{+}} \end{array} \right) g(\varphi^{t_{0}}(\theta))^{-1} \left(
\begin{array}{c} \hat{p} \\ \hat{x} \end{array} \right).
\end{array} $$
Swapping $ t $ with $ t_{0} $ in this equation and using the identity
$ U^{\dagger}(t,t_{0};\theta) = U(t_{0},t;\theta) $, we obtain
$$ \begin{array}{l}
\hspace{-0.2cm} U(t,t_{0};\theta) \, L_{\underline{\alpha}(\varphi^{t_{0}}
(\theta))} \, U^{\dagger}(t,t_{0};\theta)
= \left( \begin{array}{c} \alpha_{p}(\varphi^{t_{0}}(\theta)) \\
                     \alpha_{x}(\varphi^{t_{0}}(\theta)) \end{array} \right)^{\tra}
\left( \begin{array}{c}  U(t,t_{0};\theta) \, \hat{p} \, U^{\dagger}(t,t_{0};\theta) \\
                    U(t,t_{0};\theta) \, \hat{x} \, U^{\dagger}(t,t_{0};\theta) \end{array}
\right) \\[0.4cm]
\hspace{0.5cm} = \left( \begin{array}{c} \alpha_{p}(\varphi^{t_{0}}(\theta)) \\
                    \alpha_{x}(\varphi^{t_{0}}(\theta)) \end{array} \right)^{\tra}
g(\varphi^{t_{0}}(\theta)) \left( \begin{array}{cc} e^{-(t-t_{0})\lambda_{+}} & 0 \\
         0 & e^{(t-t_{0})\lambda_{+}}\end{array} \right) g(\varphi^{t}(\theta))^{-1}
\left( \begin{array}{c} \hat{p} \\ \hat{x} \end{array} \right).
\end{array} $$
Thus, we can deduce the stable and unstable directions
$$ \left( \begin{array}{c} \alpha_{p-}(\theta) \\ \alpha_{x-}(\theta) \end{array} \right) =
\left( g(\theta)^{-1} \right)^{\tra} \left( \begin{array}{c} 1 \\ 0 \end {array} \right)
~~~~ \text{and} ~~~~
\left( \begin{array}{c} \alpha_{p+}(\theta) \\ \alpha_{x+}(\theta) \end{array} \right) =
\left( g(\theta)^{-1} \right)^{\tra} \left( \begin{array}{c} 0 \\ 1 \end {array} \right). $$
Writing $ g=(g_{ij})_{1\leq i,j\leq 2} $, we obtain
$$ \left\lbrace \begin{array}{l} \alpha_{p-}(\theta) = g_{22}(\theta) \, \det(g(\theta))^{-1} \\
\alpha_{x-}(\theta) =  -g_{12}(\theta) \, \det(g(\theta))^{-1} \end{array} \right.
~~~ \text{and} ~~~~
\left\lbrace \begin{array}{l} \alpha_{p+}(\theta) = -g_{21}(\theta) \, \det(g(\theta))^{-1} \\
\alpha_{x+}(\theta) = g_{11}(\theta) \, \det(g(\theta))^{-1} \end{array} \right.
\hspace{-0.2cm} , $$
with
$$ U(t,t_{0};\theta) \, L_{\underline{\alpha}_{\pm}(\varphi^{t_{0}}(\theta))} \,
U^{\dagger}(t,t_{0};\theta) \, = \, e^{\pm \lambda_{+}(t-t_{0})} \,
L_{\underline{\alpha}_{\pm}(\varphi^{t}(\theta))}. $$
\cqfd

\vspace{0.4cm}

\subsection{The configurational quantum cat system}

We consider a charged particle of mass $ m=1 $ constrained to move in a unit
square with periodic boundary conditions (period 1) submitted to external periodic
time dependent electromagnetic fields. It was shown in Ref. \cite{Weig,Weig1}
that the external fields can be chosen in such a way that the configuration space
of the particle is mapped periodically to itself according to Arnold's cat map.
This system is described by the Hamiltonian
\begin{equation*}
  H(\hat{x},\hat{p},t) \, = \, \frac{1}{2}\, \hat{p}^{\tra} \, \hat{p} \, +
  \, \frac{1}{2}\, (\hat{p}^\tra \, A+A^\tra \, \hat{p})
\end{equation*}
with $ \hat{p}\, = \left( \begin{array}{c} \hat{p}_1 \\ \hat{p}_2 \end{array} \right) $
and $ \hat{x}\, = \left( \begin{array}{c} \hat{x}_1 \\ \hat{x}_2 \end{array} \right) $.\\
The vector potential $ A $ of the fields has the form
$$ A \, = \, V \, \hat{x} \, \Delta_{T,\varepsilon} \, , $$
where $ \Delta_{T,\varepsilon} $ is a sequence of smooth kicks of period $ T $ and
duration $ \varepsilon << T $, while $ V $ is a matrix such that $ \exp(V) $ is
Arnold's cat map:
$$ e^{V} \, = \, C \, = \left( \begin{array}{cc}2 & 1 \\1 & 1 \end{array} \right). $$
The time evolution operator over one period $ T $ or Floquet operator
$ U_F=U(T,0) $ becomes in the limit $ \varepsilon\rightarrow 0 $:
$$ U_{F} \, = \,  e^{-\frac{i\, T}{2} \hat{p}^2} \, e^{-\frac{i}{2}(\hat{x}^\tra \,
V^\tra \, \hat{p} \, + \, \hat{p}^\tra \, V \, \hat{x})}. $$
Since the configuration space of the system is a torus (as opposed to
$ \RR^{n} $ in the former examples), we need to slightly adapt the definitions of
Section 2. We choose as algebra $ \mcA $ of observables the $ C^* $-algebra
generated by the Weyl operators $ W (\beta, \gamma) \, = \, \exp \, [i \, (\beta^{\tra}
\, \hat{x} + \gamma^{\tra} \, \hat{p})] $ with $ \beta \in 2\pi \ZZ^2 $ and
$ \gamma \in  \RR^2 $. \\

\begin{remark}
In the definition 2 of the Anosov property, we can use derivations that are not
necessarily inner derivations, \ie we do not need to impose $ \underline{\alpha}_{x}
\in 2\pi \ZZ^2 $. Indeed, according to (\ref{eq6}), $ \delta_{\underline{\alpha}}(A)=
[ L_{\underline{\alpha}}, A ] \in \mcA $ for all
$ \underline{\alpha}=(\underline{\alpha}_{x},\underline{\alpha}_{p}) \in \RR^{2n} $.
\end{remark}

\begin{theo}
  The configurational quantum cat system satisfies quantum Anosov properties:
  There exist two stable directions $ \underline{\alpha}_{1} $ and
  $ \underline{\alpha}_{2} $,
  $$ U_{F} \; L_{\underline{\alpha}_{i}} \; U_{F}^{\dag} \, =
                       \, e^{-\lambda } \, L_{\underline{\alpha}_{i}} \qquad i=1,2 \, , $$
  and two unstable directions $ \underline{\alpha}_{3} $ and
  $ \underline{\alpha}_{4} $,
  $$ U_{F} \; L_{\underline{\alpha}_{i}} \; U_{F}^{\dag} \, =
                       \, e^{\lambda } \, L_{\underline{\alpha}_{i}} \qquad i=3,4 \, , $$
  where $ \lambda \, > \, 0 $ is such that $ e^{ \pm \lambda } $ are the eigenvalues
  of Arnold's cat map $ C $.
\end{theo}
\proo
The operator
$ D_{_V} \, = \, e^{-\frac{i}{2}(\hat{x}^\tra \, V^\tra \, \hat{p} \, + \, \hat{p}^\tra \, V
\, \hat{x})} $, one of the factors of the evolution operator $ U_{F} $, is a dilatation:
\begin{eqnarray*}
 D_{_V}^{\dag} \; \hat{x} \; D_{_V} & = & e^V \, \hat{x} ~ \, = \, ~ C \, \hat{x},  \\
 D_{_V}^{\dag} \; \hat{p} \; D_{_V} & = & e^{-V^\tra} \, \hat{p} ~ \, = \, ~
 C^{-1} \, \hat{p}.
\end{eqnarray*}
The evolution of the position and momentum operators over one period $ T $ is thus
given by
\begin{eqnarray}
 U_{F} \; \hat{x} \; U_{F}^{\dag} & = & C^{-1} \, \hat{x}
                                                             + \, T \, C^{-1} \, \hat{p} \label{UxU} \\
 U_{F} \; \hat{p} \; U_{F}^{\dag} & = & C \, \hat{p}. \label{UpU}
\end{eqnarray}
Equation (\ref{UpU}) allows one to conclude that
$$ U_{F} \; L_{(0,\alpha_p)} \; U_{F}^{\dag}
                                                    \, = \, \alpha_p^{\tra} \, C \, \hat{p} \, , $$
with $ L_{\underline{\alpha}} = L_{(\alpha_{x},\alpha_{p})} = \alpha_{x}^\tra \,
\hat{x} \, + \, \alpha_{p}^\tra \, \hat{p}  $.
Therefore  $ \underline{\alpha}_{1} = ( \, 0 \, , \, v_{-} ) $ and $ \underline{\alpha}_{3}
= ( \, 0 \, , \, v_{+} ) $ are respectively stable and unstable directions, where
$ v_{\pm} $ are the eigenvectors of $ C $ with $ C \, v_{\pm} \, =
\, e^{\pm \lambda } \, v_{\pm} $.  \\

\noindent Using equations (\ref{UxU}) and (\ref{UpU}), we observe that
\begin{eqnarray*}
 U_{F} \; [ \, ( \, C^2 - Id \, ) \, \hat{x} \, + \, T \, \hat{p} \, ] \; U_{F}^{\dag} & = &
 ( \, C^2 - Id \, ) \, C^{-1} \, ( \, \hat{x} \, - \, T \, \hat{p} \, ) \, + \, T \, C \, \hat{p} \\
 & = &  C^{-1} \,  [ \, ( \, C^2 - Id) \, \hat{x} \, + \, T \, \hat{p} \, ].
\end{eqnarray*}
Hence $ \underline{\alpha}_{2} = (\, ( \, C^2 -  Id \, ) \, v_+ \, , \, T \, v_+ ) $ and
$ \underline{\alpha}_{4} = (\, ( \, C^2 - Id \, ) \, v_- \, , \, T \, v_- ) $ are the
second pair of stable and unstable directions.  \\
\cqfd

\begin{remark}
The derivations $ \delta_{\underline{\alpha}_{1}} $ and $ \delta_{\underline{\alpha}_{3}} $
are inner derivations, but $ \delta_{\underline{\alpha}_{2}} $ and
$ \delta_{\underline{\alpha}_{4}} $ are not because the coefficients of each
eigenvector $ v_{\pm} $ are rationally independent.
\end{remark}

\begin{remark}
It follows immediately from the Anosov properties that the upper Lyapunov
exponent for this system is $ \overline{\lambda}=\lambda>0 $.
\end{remark}

\vspace{2cm}

\bibliography{refearticle}

\end{document}